**The Bottom of the Main Sequence and Beyond: Speculations, Calculations, Observations, and Discoveries (1958-2002)**

Shiv S. Kumar

*The Galileo Institute, P.O. Box 6516, Charlottesville, VA 22906, U.S.A. & Department of Astronomy, University of Virginia, Charlottesville, VA, U.S.A.*

**Abstract.**  In this paper, I briefly review the theoretical and observational work done since 1958 on the gaseous objects with mass below the H-burning limit. Special attention is paid to the theory that I developed during the period 1958-1962 for the hydrogen-rich gaseous objects with mass below the minimum main sequence mass of ~ 0.08 Msun. The three main predictions of this theory (Kumar 1962a; Kumar 1962b; Kumar 1963a; Kumar 1963b) are discussed. Fragmentation processes for interstellar clouds are discussed, and it is pointed out that the minimum mass of a gaseous fragment may be as low as 0.001 Msun. Observational results obtained since 1995 on the luminous and dark objects of very low mass are briefly reviewed. Comments are made on the basic nature of the stars and planets, and it is pointed out that the processes of star formation are fundamentally different from those of planet formation. Arguments are presented to show that some of the very-low-mass dark companions, discovered since 1995, were most probably formed by the star formation processes and not by the planet formation processes.

## 1. Introduction

I started working on the effects of electron degeneracy on the structure and evolution of the contracting stars of very low mass in the Fall of 1958 at the University of Michigan (Ann Arbor). Until then, the theoretical calculations on the gravitational contraction phase (pre-main sequence phase) of stellar evolution (Levee 1953; Henyey et al. 1955) were carried out under the assumption that the material in the interiors of the contracting stars was in the form of perfect (ideal) gas. However, since the central temperature of a contracting star varies as $M/R$ and its central density as $M/R^3$, where M and R are the mass and radius of the star, respectively, it became clear to me that, for a sufficiently low mass, the slow rise in the central temperature coupled with the rapid rise in the central density will cause the gaseous material to become partially degenerate before the central temperature could get high enough for the onset of the hydrogen burning thermonuclear reactions. Preliminary calculations based on the interior models in radiative equilibrium, approximated by polytropes of index 3.0 (Levee 1953), convinced me that, for a star of mass ~0.15 Msun, the electron degeneracy effects would become important at a central temperature of a few million Kelvins. Any further contraction of the star would then cause the central temperature to go through a maximum value that was less than that required for the onset of the proton-proton chain reactions.

Although my calculations at the time (1958-1959) were quite preliminary, I was convinced that below a certain critical mass the star would not be able to become a main–sequence object. But that conclusion immediately led me to ask the following two questions: 1. Can star formation processes actually give rise to the gaseous objects with mass below the minimum H-burning mass? and 2. What does eventually happen to these objects, if they exist? In order to answer the first question, I had to study the star formation problem, which in those days was done in terms of the Jeans criterion. Although the Jeans criterion didn't help me in finding the answer to my question, I did manage to obtain a modification of the criterion when the effects of viscosity and thermal conductivity were included in the study of the gravitational instability problem (Kato & Kumar 1960; Kumar 1960). After struggling with this question for a year or so, I finally came to the conclusion that there really was no good reason for the star formation processes not to give birth to the gaseous objects with mass below the minimum H-burning mass. Intuitively, I felt very strongly that the star formation processes should go on beyond and below the minimum H-burning mass limit since these processes had no foreknowledge of the existence of this critical mass.

Next I attempted to answer the second question raised above. The qualitative answer that I came up with was that the star would keep on collapsing beyond the stage of maximum central temperature. As a result of the compression of the partially degenerate electrons in the interior of the star, the temperature would start decreasing and this in turn will take the star closer to the state of complete electron degeneracy. The collapse process would finally come to an end when the material inside the star becomes a completely degenerate object.

During my stay at the Smithsonian Astrophysical Observatory (Cambridge) in 1960-1961, I continued my work on very low mass stars. After discussing my results with David Layzer, Charles Whitney and a few other people, I wrote a paper dealing with my ideas on the possible existence of the gaseous objects with mass below the H-burning limit. This paper was submitted to the journal PASP in July 1961, but the paper was not accepted for publication, since the referee did not agree with my conclusions. Interestingly enough, even though the referee didn't think that my conclusions were correct, he did say that my results would be of great importance if they turn out to be correct!

On February 6, 1962, I joined the NASA Institute for Space Studies (New York) as NAS-NRC Postdoctoral Research Associate (approximately six weeks before my 23$^{rd}$ birthday) and resumed my calculations on the properties of the very-low-mass gaseous objects that, despite the negative comments of the PASP referee, I was sure, couldn't and wouldn't go through the p-p chain reactions. However, in addition to using interior models in radiative equilibrium (polytropes of index 3.0), I used interior models in convective equilibrium (polytropes of index 1.5) after seeing a copy of Hayashi's paper (Hayashi 1961) in which he had suggested that the contracting stars of low mass might be in con-vective equilibrium. (It should be noted here that the lowest mass considered by him was 0.4 Msun). But the change from the completely radiative interiors to the completely convective interiors had no effect on my earlier conclusion regarding the paramount importance of the electron degeneracy effects for the evolution of very low mass stars. Whether in convective equilibrium or in radiative equilibrium, my conclusion still was that the stars with mass below a certain critical mass (the minimum H-burning mass) would keep on contracting until they become completely degenerate objects.

In April 1962, I submitted another paper on the structure and evolution of the gaseous objects with mass below the H-burning limit, which by then I had estimated to have a numerical value of ~ 0.1 Msun, to the PASP. I was very optimistic that this time around the paper would be accepted for publication, but in July 1962 I received a letter of rejection from the editor. I was informed that the referee (and presumably the editor) didn't agree with my conclusions, and therefore my paper couldn't be published in the PASP.

## 2.  The Bottom of the Main Sequence and Beyond: Theoretical Predictions

From April to July (1962), I worked on the convective interior models for stars with the following masses: 0.09 Msun, 0.08 Msun, 0.07 Msun, 0.06 Msun, 0.05 Msun and 0.04 Msun. For all of these masses, the internal structure was computed by making use of the correct equation of state incorporating the effects of non-relativistic electron degeneracy. Models were computed at various radii for objects in the above-mentioned mass range with Population I and Population II chemical compositions. My views on the star formation processes, and the application of these models to the contracting stars allowed me to make the following theoretical predictions in July 1962:
1. Star formation processes should produce gaseous objects with mass below the H-burning limit, which I found to have a numerical value of approximately 0.07 Msun for Population I stars and approximately 0.09 Msun for Population II stars.
2. Over a time scale much shorter than the age of the Galaxy, the gaseous objects with mass below the H-burning limit become completely degenerate, low-luminosity objects.
3. The Galaxy is likely to contain a very large population of the hydrogen-rich and intrinsically very faint objects of very low mass that are composed of degenerate matter.

      I presented a paper giving the details of my theoretical calculations as well as the predictions of the new theory at the 111$^{th}$ meeting of the American Astronomical Society held at Yale University in August 1962. The abstract of this paper was published in the Astronomical Journal in November 1962 (Kumar 1962b). I also submitted two papers describing the new theory to the Astrophysical Journal in October 1962. These papers were quickly accepted for publication in November 1962 and published in May 1963 (Kumar 1963a; Kumar 1963b).

Since 1963, many workers have carried out detailed numerical calculations on the structure and evolution of the gaseous objects with mass below the H-burning limit (referred to as the Kumar limit by scientists such as P. Demarque, J. Hills, and J. Patterson) which have confirmed the above-mentioned results (see, for example, Grossman 1970; Nelson et al. 1985; D'Antona & Mazzitelli 1985). See also the recent review articles by Basri (2000), Chabrier & Baraffe (2000), and Oppenheimer et al. (2000) and the book by Reid & Hawley (2000) for additional references. Observational studies confirming the theoretical predictions will be discussed in section 4 of this paper.

## 3.  Star Formation Processes and the Minimum Fragmentation Mass

In late 1962, I started to extend my calculations to objects with mass lower than 0.04 Msun. The treatment of the equation of state for these objects was problematic. An even more serious (and more difficult) problem was to figure out the final mass limit for a gaseous fragment formed by the star formation processes. Although I felt that the minimum fragment mass could be lower than 0.01 Msun, for practical reasons I terminated the computations of the interior models at a mass of 0.01 Msun. My results on the structure and evolution of a star of mass 0.01 Msun were initially submitted to the Astronomical Journal in 1963, but the then-editor of the journal returned the paper saying that the Astronomical Journal was not the proper journal for my paper!

This paper was eventually published in Zeitschrift fur Astrophysik (Kumar 1964). It was in this paper that I pointed out, for the first time, that the end-product of the evolution of a gaseous object with mass below the H-burning limit is **not** a planet. The nature of the very-low-mass dark companions, which had reportedly been detected until 1963 by the astrometric method, was also discussed in this paper. It was pointed out that the dark companions, with their reported masses and orbital charateristics, most probably originated by the star formation processes. The arguments used in 1963-1964 to conclude that the dark companions with a reported mass range of 0.0015-0.02 Msun are not likely to be planets are still valid today, and they will be discussed further in sections 4 and 5 of this paper.

My conclusion that the minimum mass of a gaseous fragment, formed by the star formation processes, is much smaller than the minimum H-burning mass was discussed further in a series of papers published during the period 1967-2000 (Kumar 1967; Kumar 1972b; Kumar 1990; Kumar 1994; Kumar 1995; Kumar 2000b). My current position is that the fragmentation of the interstellar (and primordial) clouds, in general, should give rise to gaseous fragments with mass as low as 0.001 Msun. The basic argument is that, if a just-formed gaseous fragment with a temperature ~ 10 K can achieve a mean density of ~ $1 \times 10^{-13}$ gm cm$^{-3}$ , then it'll be able to collapse under its own gravity as an independent object (Kumar 1990; Kumar 2000b). It appears that the gravitational instability in the collapsing cores of the molecular clouds may be able to give rise to fragments with such high densities.

In the past three decades, a few other workers (see, for example, Low & Lynden-Bell 1976; Silk 1977; Boss 1987; Boss 2001) have also come to the conclusion hat the fragmentation of interstellar clouds can give rise to gaseous fragments with mass much lower than the H-burning limit. Low and Lynden-Bell estimated the minimum fragmentation mass to be approximately 0.007 Msun, while Silk proposed, for spherically collapsing clouds, a numerical value of ~ 0.01 Msun. Boss has also proposed similar values for the minimum fragmentation mass.

**4.   The Luminous and Dark Objects of Very Low Mass: Observational Results**

During 1962-1963, there existed some doubt in the astronomical community concerning the existence of the gaseous objects with mass below the H-burning limit (that I had just postulated), but observational astronomers such as Victor Blanco, Olin Eggen, and Jesse Greenstein started to search for them soon after they learned the details of my theoretical results. (During 1962-1963, George Herbig had also shown interest in searching for them). Eggen actually started searching for them in 1962 and seven years later reported on their possible detection (Eggen 1969). Greenstein also discussed the relevant observational data concerning their existence in 1969 (Greenstein 1969). Greenstein referred to the gaseous objects with mass below ~ 0.08 Msun as the Kumar objects, as did many other workers during the period 1962-1990 before the terminology of 'brown dwarfs' became popular. In the rest of the paper, the gaseous objects with mass below the H-burning limit, formed by the star formation processes, will be referred to as brown dwarfs.

   From the observational point of view, not much real progress was made in this field until 1995. During 1995, the first convincing observational evidence concerning the existence of brown dwarfs was obtained (Rebolo et al. 1995; Nakajima et al. 1995; Oppenheimer et al. 1995). Rebolo et al. found an object, which they called Teide 1, in Pleiades that had the expected characteristics of an object with mass below the Kumar limit. Similarly, Nakajima et al. and Oppenheimer et al. discovered and studied a faint, cool companion to the star Gliese 229 that had the expected properties of a brown dwarf.

    Before discussing some of the observational results obtained since 1995, I want to briefly describe the three important stages in the evolution of the objects with mass below the Kumar limit. The first stage is the initial gravitational collapse phase during which the object moves more or less vertically downward in the H-R diagram. The color of the object is red (or very red) during this stage.

   The object enters its second evolutionary stage when the gaseous material in the interior becomes completely degenerate, and the object then reaches its cooling curve in the H-R diagram. For a given mass (and chemical composition), the cooling curve can be constructed using the radius corresponding to the stage of complete electron degeneracy (Kumar 1962a; Kumar 1963a). During this stage the object, still luminous, starts sliding down on its cooling curve, and its color gets redder and redder as the surface temperature decreases with time. I would like to designate here the collection of all the cooling curves (in the H-R diagram) for the hydrogen-rich degenerate objects with mass below the H-burning limit as **the NHB Sequence**. The NHB Sequence defines a region (zone) in the H-R diagram for the hydrogen-rich cooling degenerate objects (of very low mass) just as the Main Sequence defines a zone in the H-R diagram for the hydrogen burning (HB) objects. And since the contracting stars with mass **above** the H-burning limit are referred to as the pre-main-sequence objects, we should refer to the contracting stars with mass **below** the H-burning limit (in the first evolutionary stage) as the pre-NHB-sequence objects. Consequently, when talking about the contracting very-low-mass objects, phrases such as 'pre-main-sequence brown dwarfs' should be avoided. If brown dwarfs are defined as the objects that are unable to go through the main sequence evolutionary stage, then it makes no sense to talk of 'pre-main-sequence brown dwarfs.'

The third evolutionary stage for the very-low-mass objects is reached when their luminosities become so small that we cannot detect them as luminous objects. This third (and final) stage of evolution was referred to as the 'black dwarf' stage in my 1962-1963 papers. During this stage, the objects may alternatively be referred to as the 'dark degenerate dwarfs' or as 'dark brown dwarfs', but I still think that the terminology of 'black dwarfs' is the right choice for the third stage. The important thing to keep in mind is that we have to separate the second stage, when the objects are luminous, from the third stage, when they are dark.

Since 1995, hundreds of papers have been published in which new and exciting observational results have been reported concerning the properties of the very-low-mass objects in various evolutionary stages. In the following paragraphs, I'll briefly discuss some of these results. I regret that, due to space limitations, I won't be able to discuss all the exciting observational results that have been obtained by so many hard-working scientists during the past seven years.

I find the reported discovery of many very-low-mass luminous objects in Orion (Zapatero Osorio et al. 2000; Lucas & Roche 2000; Lucas et al. 2001; Bejar et al. 2001; Martin et al. 2001; Zapatero Osorio et al. 2002) to be very exciting. From their observed locations in the H-R diagram, these objects appear to be very-low-mass objects (brown dwarfs) in the pre-NHB-sequence stage of evolution. The really exciting thing is that some of these objects may have masses of ~ 0.005 $M_{sun}$ (or ~ 5 $M_{jup}$). To me, the discovery of these objects represents a confirmation of the basic theory of the very-low-mass gaseous objects (discussed earlier in sections 1-3). The probable detection of an isolated (single) object with an estimated mass of a few Jupiter masses (Zapatero Osorio et al. 2002) indicates that the fragmentation of interstellar clouds can produce such objects as predicted by the theoretical investigations discussed in sections 1-3.

The exciting (but not surprising) discovery of the circumstellar disks around some very-low-mass young objects (the pre-NHB-sequence objects) by a number of workers (see, for example, Natta & Testi 2001; Muench et al. 2001; Apai et al 2002) also provides support for the basic theory of the origin and evolution of brown dwarfs. In particular, the presence of circumstellar disks around the pre-NHB-objects shows clearly that the star formation processes don't have any foreknowledge about the H-burning limit. If young pre-main-sequence stars, such as T Tauri stars, can be accompanied by circumstellar disks, then we should expect to see such disks around young brown dwarfs as well. Whether circumstellar disks around young brown dwarfs will later give birth to planetary systems remains to be seen. In general, I expect the environment for the formation of planets in the vicinity of brown dwarf stars to be more favorable than that in the vicinity of stars with mass ~ 1.0 Msun.

The existence of the luminous objects with mass below the H-burning limit in double and multiple stars also provides observational support for the idea that brown dwarfs, like the stars with mass above the H-burning limit, are formed by the star formation processes. Since the discovery of Gliese 229B, several brown dwarfs have been detected as members of double and multiple systems (see, for example, Kirkpatrick et al. 2001; Kenworthy et al. 2001; Close et al. 2002; Goto et al. 2002). These observations clearly show that all members in the double and multiple systems, including the members with mass below ~ 0.08 Msun, originated by a common (star) formation process.

Next I want to make a few comments on the nature of the recently discovered **dark** companions with mass below ~ 0.08 Msun. I would like to point out once again that some, and possibly most, of these companions are not likely to be planets. Since 1995, quite a few dark companions (in double or multiple systems) with mass in the range 0.001-0.02 Msun have been detected by measuring the radial velocity variations in the spectra of the primary stars. These dark objects (secondaries) are being referred to as 'extrasolar planets' by many people in the scientific community, but I think that it's much more likely that some of them are hydrogen-rich degenerate objects of very low mass in the third (and final) stage of their evolution. From the measured (minimum) masses of the companions **and** their orbital characteristics, it is clear to me that some of these systems originated as double or multiple stars. Take, for example, the case of the dark companion to the star HD 136118 (Fischer et al. 2002). The companion has a minimum mass of 11.9 Mjup (~ 0.012 Msun), an orbital eccentricity of 0.37 and a semi-major axis of 2.3 AU. With these properties, it appears to me that the visible star and its dark companion most probably originated as a double star system. Other systems, such as HD 106252 and HD 50554 (Fischer et al. 2002), are very likely to have originated as double stars and not as star-planet systems.

Finally, let me acknowledge the efforts of the workers involved in 2MASS, SDSS, and DENIS. I'm delighted to see that these surveys have provided us with sufficient evidence to conclude that the Galaxy contains a very large population (billions and billions!) of the hydrogen-rich objects with mass below the H-burning limit as predicted back in 1962-1963 (section 2).

## 5.   The Basic Nature of the Stars and Planets

Since 1964, I have been presenting arguments to show that the star formation processes are fundamentally different from those of planet formation (Kumar 1964; Kumar 1967; Kumar 1972a; Kumar 1974; Kumar 1990; Kumar 1995; Kumar 2000a). Stars (including brown dwarfs) are formed by the fragmentation of gaseous clouds, and the mass range in the stellar domain ranges from a few hundred solar masses to ~ 0.001 Msun or ~ 1 Mjup. Planets are formed by the slow accumulation (accretion) of dust, rocks, and gas in the vicinity of a star, and the mass range in the planetary domain ranges from ~ 0.000001 Mjup to ~ 2 Mjup. Thus, as far as the masses of the stars and planets are concerned, I'm not talking of just one linear sequence but of two separate sequences arising from two different formation processes.

The old idea that the stars and planets represent two sections of the same linear sequence (in mass), with objects above a certain mass labeled as stars and objects below that certain mass labeled as planets, is not the correct way to understand the two groups of objects. The mass of an object doesn't uniquely determine its basic nature. In order to ascertain the basic nature of an object, we have to know its formation mechanism. Theoretically speaking, an object of 1 Mjup (located somewhere in the Universe) may come into existence by either the star formation processes or the planet formation processes. As I have repeatedly pointed out (Kumar 1972a; Kumar 1974; Kumar 1994; Kumar 1995), the most massive planet in the Solar System (Jupiter) was most probably formed by the planet formation processes and not by the star formation processes. For Jupiter, the presence of a rocky/metallic core, the chemical composition of its interior, and the chemical composition of its atmosphere clearly indicate that the planet acquired its present mass (in the presence of the Sun) by the slow accretion of dust, rocks, and gas over the past 4.5 billion years (Kumar 1994; Kumar 1995; Kumar 2000a). It does not appear to have been formed by the rapid collapse of an extended, gaseous object of mass 0.001 Msun.

I thank Dr. Eduardo Martin and the SOC of the IAU Symposium #211 for inviting me to present this paper and to be a panelist in the panel discussion at the meeting at Waikoloa, Hawaii.

## References

Apai, D., et al. 2002, ApJ, 573, L115
Bejar, V. J. S., et al. 2001, ApJ, 556, 830
Basri, G. 2000, ARA&A, 38, 485
Boss, A.P. 1987, ApJ, 319, 149
Boss, A.P. 2001, ApJ, 551, L167
Chabrier, G. & Baraffe, I. 2000, ARA&A, 38, 337
Close, L. M. et al. 2002, ApJ, 567, L53
D'Antona, F. & Mazzitelli, I. ApJ, 1985, 296, 502
Eggen, O.J. 1969, ApJ, 157, 287
Fischer, D. et al. 2002, PASP, 114, 529
Goto, M. et al. 2002, 567, L59
Greenstein, J. L. 1969, in *Stellar Astronomy*, vol. 2, ed. H. Y. Chiu et al., Gordon & Breach Science Publishers
Grossman, A. S. 1970, ApJ, 161, 619
Hayashi, C. 1961, PASJ, 13, 450
Henyey, L. G. 1955, PASP, 67, 154

Kato, S. & Kumar, S. S. 1960, PASJ, 12, 290
Kenworthy, M. et al. 2001, ApJ, 554, L67
Kirkpatrick, J.D. et al. 2001, AJ, 121, 3235
Kumar, S. S. 1960, PASJ, 12, 552
Kumar, S. S. 1962a, Institute for Space Studies Report no. x-644-62-78
Kumar, S. S. 1962b, AJ, 67, 579
Kumar, S. S. 1963a, ApJ, 137, 1121
Kumar, S. S. 1963b, ApJ, 137, 1126
Kumar, S. S. 1964, ZfAp, 58, 248
Kumar, S. S. 1967, Icarus, 6, 136
Kumar, S. S. 1972a, Ap&SS, 16, 52
Kumar, S. S. 1972b, Ap&SS, 17, 219
Kumar, S. S. 1974, Ap&SS, 28,173
Kumar, S. S. 1990, ComAp, 15, 55
Kumar, S. S. 1994, Ap&SS, 212, 57
Kumar, S. S. 1995, ASP Conference Series, 74, 231
Kumar, S. S. 2000a, Publ. Galileo Inst., Vol. 1, No. 1
Kumar, S. S. 2000b, Publ. Galileo Inst., Vol. 1, No. 2
Levee, R. D. 1953, ApJ, 117, 200
Low, C. & Lynden-Bell, D. 1976, MNRAS, 176, 367
Lucas, P.W. & Roche, P. F. 2000, MNRAS, 314, 558
Lucas, P. W. et al. 2001, MNRAS, 326, 695
Martin, E. L. et al. 2001, ApJ, 558, L117
Muench, A.A. et al. 2001, ApJ, 558, L51
Nakajima, T. et al. 1995, Nature, 378, 463
Natta, A. Testi, L. 2001, A&A, 376, L22
Nelson, L. A. et al. 1985, Nature, 316, 42
Oppenheimer, B. R. et al. 1995, Science, 270, 1478
Oppenheimer, B. R. et al. 2000, in *Protostars and Planets IV*, ed. Mannings, V. et al., Tucson: University of Arizona Press
Reid, I. N. & Hawley, S. L. 2000, *New Light on Dark Stars*, Springer
Rebolo, R. et al. 1995, Nature, 377, 129
Silk, J. 1977, ApJ, 214, 152
Zapatero Osorio, M. R. et al. 2000, Science, 290, 103
Zapatero Osorio, M. R. et al. 2002, ApJ, (in press)